\documentclass[aps,prb,preprint,groupedaddress,showpacs]{revtex4}

\usepackage{graphicx}
\usepackage{dcolumn}
\usepackage{bm}

\begin{document}


\title{Anisotropy of upper critical field in one-dimensional organic system, (TMTTF)$_2$PF$_6$ under extremely high pressure}


\author{Mika Kano}
\email[]{kano@magnet.fsu.edu}
\affiliation{National High Magnetic Field Laboratory, Tallahassee, FL 32301, USA}
\affiliation{School of Science, the University of Tokyo, Hongo, Tokyo 113-0033, Japan}

\author{Hatsumi Mori}
\affiliation{Institute for Solid State Physics,  the University of Tokyo, Kashiwa-shi, Chiba 277-8581, Japan}

\author{Toshikazu Nakamura}
\affiliation{Institute for Molecular Science, Okazaki 444-8585, Japan}

\author{Kazuyuki Matsubayashi}
\affiliation{Institute for Solid State Physics, the University of Tokyo, Kashiwa-shi, Chiba 277-8581, Japan}

\author{Masato Hedo}
\affiliation{Faculty of Science, University of the Ryukyus, Nishihara, Okinawa, 903-0213, Japan}

\author{Miho Itoi}
\affiliation{Department of Physics and Mathematics, College of Science and Engineering, Aoyama Gakuin University, Sagamihara, Kanagawa 229-8558, Japan}

\author{Tim Murphy}
\affiliation{National High Magnetic Field Laboratory, Tallahassee, FL 32301, USA}

\author{Stanley W. Tozer}
\affiliation{National High Magnetic Field Laboratory, Tallahassee, FL 32301, USA}

\author{Yoshiya Uwatoko}
\affiliation{Institute for Solid State Physics, the University of Tokyo, Kashiwa-shi, Chiba 277-8581, Japan}


\date{\today}

\begin{abstract}

We have measured the temperature dependent  resistivity of (TMTTF)$_2$PF$_6$ to 7 GPa using a turnbuckle DAC (diamond anvil cell)  and in magnetic field up to 5 T. Unlike many other organic conductors, a zero resistance was observed in the superconducting state even under high pressures. Superconductivity was observed over a range of $P$ = 4.18 GPa to 6.03 GPa and showed a peak $T_c$ of 2.25 K at 4.58 GPa. The temperature dependence of the upper critical magnetic field $H_{c2}(T)$ was determined via resistivity at $P$ = 4.58 GPa, for the intrachain ($a$), interchain ($b'$), and interlayer ($c^*$) configurations and the $H_{c2}(T)$ displays positive curvature without saturation, which may be originated by a FFLO state, for magnetic field along $a$-axis and $b'$-axis in T $\ge$ 0.5 K for $P$ = 4.48 GPa. This feature is suppressed with increasing pressure and the orbital pair breaking mechanism becomes dominant. The values of the Ginzberg-Landau coherence length for three different axes obtained from this work shows that (TMTTF)$_2$PF$_6$ is an anisotropic three - dimensional superconductor.
\end{abstract}

\pacs{74.25.Dw, 74.25.Fy, 74.62.-c, 74.62.Fj, 74.70.Kn}

\maketitle

\section{Introduction}

The Bechgaard salt (TMTSF)$_2$PF$_6$ (TMTSF = tetramethyltetraselenafulvalene) was the first organic superconductor discovered in 1980 with a superconducting temperature of $T_c$ = 0.9 K (critical pressure, $P_c$ = 1.2 GPa). \cite{jerome:superconductivity:1980}Since this discovery, materials based on TMTSF and its derivatives, such as TMTTF (tetramethyltetrathiofulvalene), the so-called (TMT$C$F)$_2$X (X = monovalent anion) series, have been investigated extensively to search for more superconductors. \cite{seo:toward:2004,mori:materials:2006,itoi:pressure-induced:2007,itoi:anomalously:2008}They have also attracted attention for many years due to their rich  properties such as spin-Peierls (SP), charge-ordering, spin density wave (SDW) and possible unconventional superconductivity (SC)  that appear by changing counter anion or applying external pressures. \cite{ishiguro:organic:2001} Many compounds with TMTSF molecules exhibit a SC state at low pressures around 1 GPa\cite{ishiguro:organic:2001, jerome:physics:1991} whereas most compounds with TMTTF exhibit a SC transition at ultra-high pressures, for example $P_c$ = 5 GPa in (TMTTF)$_2$PF$_6$.\cite{adachi:superconducting:2000,jaccard:spin:2001} Taking advantage of low pressure induced\cite{lee:anisotropy:1997} and ambient pressure\cite{murata:upper:1987,oh:magnetic:2004} superconductivity, the studies especially on the SC phase and its upper critical field have been done extensively with TMTSF salts, but not many with TMTTF salts. 

A generic phase diagram was obtained by studies of many different materials among the - TMT$C$F series at ambient and high pressure, combining the trends seen with chemical pressure and external pressure.\cite{jerome:physics:1991, mori:materials:2006, itoi:pressure-induced:2007} The (chemical) pressure that each material displays on the phase diagram shows degrees of three-dimensionality caused by many factors such as size of anion, distance between molecule planes, distance between Selenium/Sulfur atoms and Coulomb repulsion between conducting electrons. However, for layered organic material like TMT$C$F salts, the $a$-axis direction is most compressive where TMT$C$F planes have weak van der Waals bonding due to overlapping $\pi$ orbits and it is least compressive along the $c$-axis direction where the TMT$C$F molecule and anion have ion bonding even though the applied external pressure is highly hydrostatic. In the case of (TMTSF)$_2$PF$_6$, it is compressed by 0.5 \%/kbar along $a$-axis while the normal directions are an order of magnitude less compressive\cite{morosin:compressibilities:1982} when 3 kbar (0.3 GPa) of hydrostatic pressure is applied. Therefore, even with the hydrostatic pressures, electronic structure changes anisotropically, the dimensionality of the system changes as well, and degrees of the change vary depending on the size of the anion and type of cation. It is an open question if we can treat chemical pressures and external pressures in exactly the same manner. So, by changing external physical pressure on different designer molecules to gather information, one hopes to achieve a universal understanding of quasi 1- dimensional organic conductors, (TMT$C$F)$_2$X series which might also be relevant to other low dimensional system.

Due to the technical difficulty of applying hydrostatic pressures to fragile organic compounds at low temperatures and magnetic fields to study the superconducting state, the anisotropy of the upper critical field ($H_{c2}$) in TMTTF salts had not been determined prior to this work. In the present work, the anisotropy of $H_{c2}$ in  (TMTTF)$_2$PF$_6$ was determined via resistivity at temperatures as low as $T$ = 0.5 K and in magnetic fields up to 5 T which also complement previously reported work. \cite{jaccard:spin:2001,araki:electrical:2007}

\section{experimental}

Single crystals of (TMTTF)$_2$PF$_6$ were synthesised by the electrochemical method\cite{bechgaard:properties:1980}. The dimensions of the crystal selected for resistivity measurements was $\sim$150$\times$30$\times$20$ \mu$m$^3$. The electric resistivity was measured by the standard four-probe method with AC current ($I$ = 1$\sim$50 $\mu$A) parallel to $a$-axis (intra chain).

This measurement was carried out in a $^3$He refrigerator with a 10 T magnet. Pressure was generated by using a turnbuckle type diamond anvil cell (DAC). The diamond culets are 0.9 mm in diameter, which are separated by a gasket indented to 65 $\mu$m that has a 300 $\mu$m hole as a sample chamber. Details of this pressure apparatus are discussed elsewhere. \cite{kano:electrical:2007} Graphite paint was used to make electrical contacts between the leads of pressed gold wires (10 $\mu$m in diameter) and the sample. The sample was placed inside a cavity with a liquid pressure medium, glycerin which was proven to transmit nearly hydrostatic pressure up to 7 GPa\cite{kakurai:feasibility:2008}, in the gasket which is placed in between two diamonds. The pressure is calibrated at low temperatures by monitoring the shift in the fluorescence of a ruby chip located in the cavity near the sample at $T$ = 77 K.\cite{piermarini:calibration:1975} The DAC used in this study tends to gain pressure as it is cooled down from room temperature to below liquid nitrogen temperature where the pressure stabilizes. During the measurements for $H//b'$ the probe had to be warmed up to above liquid nitrogen temperature a few times, with the result being different normal state resistive values. 

It is extremely difficult to calculate the absolute value of resistivity from transport measurements using a DAC on organic materials. This is due to the amount of conducting paint which is relatively large for the small sample and inaccurate determinations of the distance between contacts and the cross section area of the sample. Therefore we used the data from resistivity measurements of the same crystal using a Cubic Anvil Press for pressurization for normalizing our data at room temperature.\cite{araki:electrical:2007}

The magnetic field was applied along $a$, $b'$ and $c^*$ axis and the temperature dependence of the resistivity was measured over the pressure range of $P$ = 4.37 - 5.64 GPa at low temperatures. The orientation of the sample was adjusted by changing the angles of the DAC with respect to the probe when mounting. 

\section{results and discussion}
\subsection{Resistivity}
\begin{figure}
\includegraphics[width=4in]{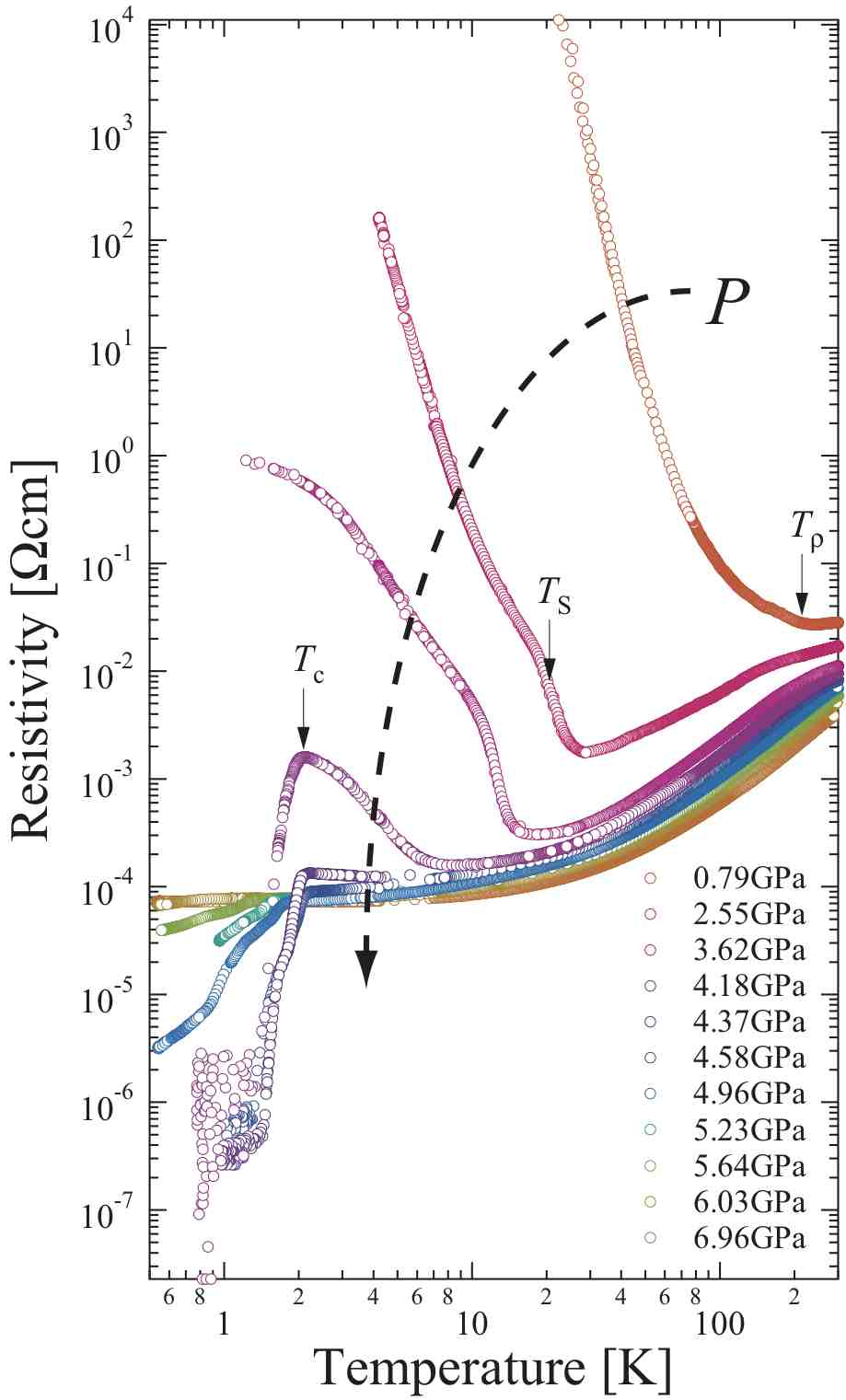}
\caption{\label{fig1}The temperature dependence of resistivity of (TMTTF)$_2$PF$_6$ in a log-log plot for $P$ = 0.79, 2.55, 3.62, 4.18, 4.37, 4.58, 4.96, 5.23, 5.64, 6.03, 6.06 GPa. $T_\rho$, $T_s$ and $T_c$ indicate transition temperatures for CL (charge localized), SDW (spin density wave) and SC (superconductivity), respectively.}
\end{figure}

The temperature dependence of resistivity of (TMTTF)$_2$PF$_6$ in a log-log plot for 0.79 $\le$ P $\le$ 6.96 GPa is shown in FIG.\ref{fig1}. 
The local minimum in resistivity observed around 240 K for $P$ = 0.79 GPa disappears by applying pressure. The  insulating behavior is gradually suppressed at higher pressure, and a metallic temperature dependence appears in the high-temperature range followed by the low-temperature insulating behavior above $P$ = 2.55 GPa. The metal insulator (M-I) transition temperature shifts to lower values as pressure is increased and it is down to 8.6 K for $P$ = 4.18 GPa. The resistivity in the insulating region shows a sharp drop mediated by the SC transition at around 2 K for 4.18 GPa. The highest SC transition is $T_c$ = 2.25 K at 4.58 GPa and the SC transition was suppressed above $P$ = 6.96 GPa. We define the local minimum temperature at 0.79 GPa as $T_\rho$, and estimated the value of $T_s$ from the local minimum of $\partial ln\rho (T)/\partial T$ where $\rho (T)$ is temperature dependent resistivity. 

\begin{figure}
   \includegraphics[width=3in]{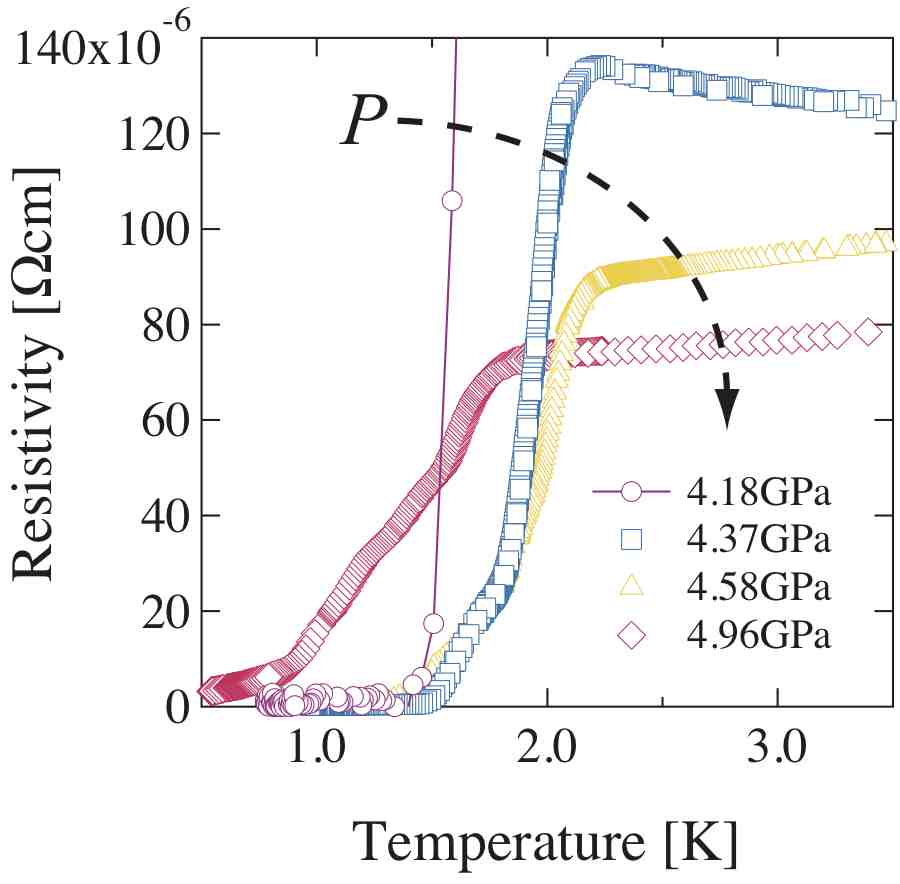}
   \caption{\label{fig2}Low temperature part of temperature dependent resistivity in linear scale at $P$ = 4.18, 4.37, 4.58, and 4.96 GPa}
\end{figure}

$T_c$ was defined by the onset temperature of the sharp drop (see arrows in FIG.\ref{fig1}). 
FIG.\ref{fig2} highlights the low temperature SC transition. Within the temperature range of this work, the presence of zero resistance was confirmed at around 1.5 K in the pressure range of 4.18 $\le$ P $\le$ 4.58 GPa which was not observed in other works.\cite{jaccard:spin:2001,adachi:superconducting:2000}
This shows that the pressures generated in present work are as hydrostatic as those of a cubic anvil press which was the only  pressure apparatus that allowed zero resistance to be observed in this material. \cite{araki:electrical:2007} The finite width of the SC transition is considered a fluctuation effect in a low dimensional system. The fluctuation effect associated with transitions cannot be ignored in low dimensional systems, the superconductivity is destroyed in finite temperature and it produces a finite resistance. This resistance decreases exponentially as the temperature drops, so it is obviously shown right below the $T_c$. This is considered as the origin of the broad transition in our case, and this behavior is typical of organic superconductors. With increased pressure, even broader transitions are observed, probably due to suppressions of spin fluctuations which is discussed later in this section.

Hump structures seen below T$_c$ have not been observed in earlier works. \cite{adachi:superconducting:2000,jaccard:spin:2001,araki:electrical:2007} We consider that these are attributed to an inhomogeneous SC state which may be due to microcracks related to the extreme sample brittleness, and not intrinsic property.

\begin{figure}
   \includegraphics[width=3in]{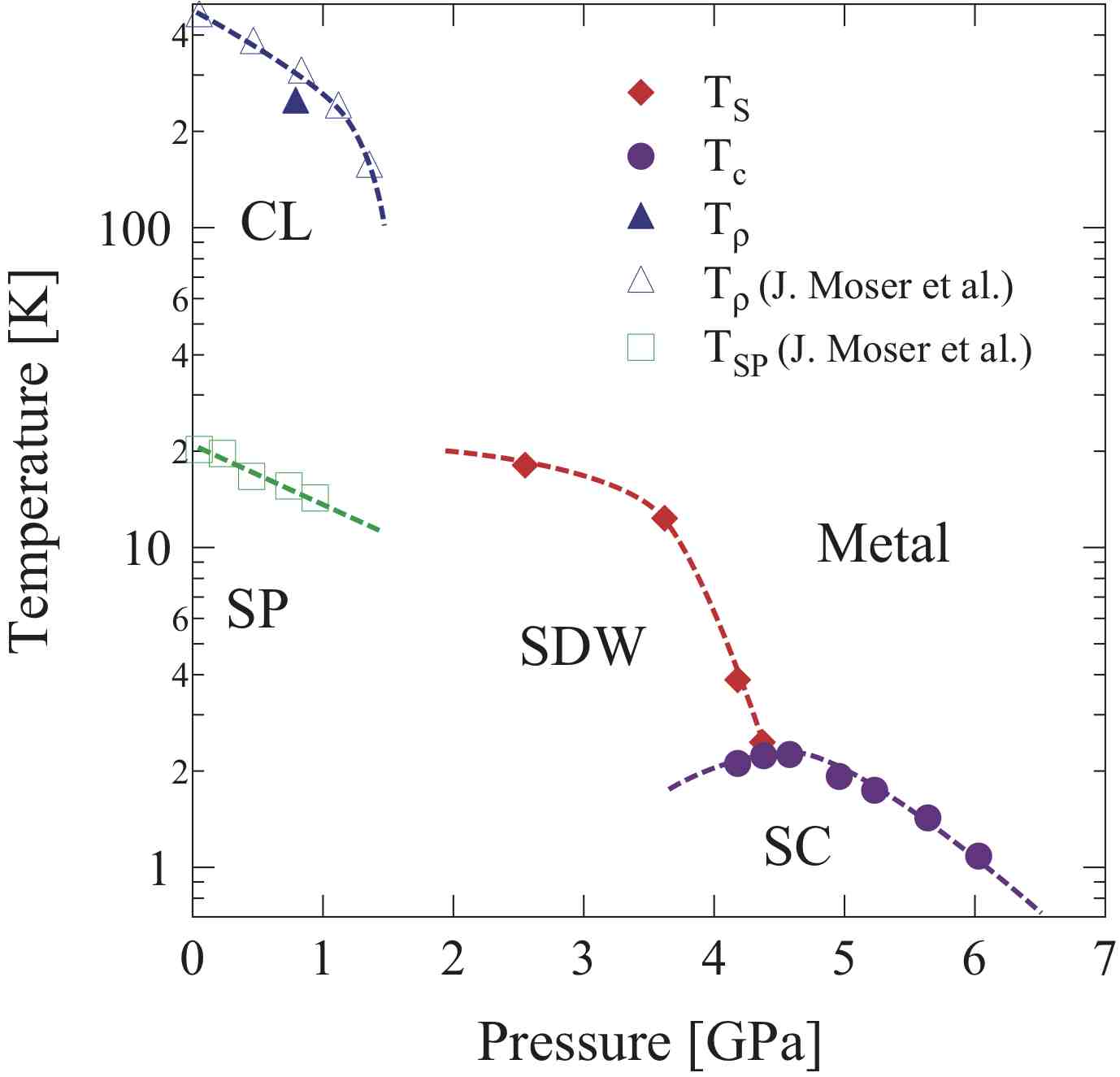} 
   \caption{\label{fig3}$P-T$ phase diagram obtained from this work. $T_s$, $T_c$, $T_\rho$, $T_{SP}$ represents transition temperatures of SDW, SC, CL, and SP respectively. Opened symbols represent data taken from ref.\onlinecite{moser:transverse:1998}}
\end{figure}

In the generic $P-T$ (pressure-temperature) phase diagram\cite{mori:materials:2006, itoi:pressure-induced:2007, dumm:mid-infrared:2005}, the ground state of (TMTTF)$_2$PF$_6$ becomes an AF/SDW from SP at around 1 GPa then becomes a SC phase at around 4.3 GPa. The local minimum of the resistance appears at around 250 K at ambient pressure, and this is known as the charge localised (CL) or Mott transition. Using these values, $T_\rho$ and $T_s$ are defined by the gap formation of CL and SDW, respectively.
FIG.\ref{fig3} is the $P-T$ phase diagram obtained from this work. 
Note that open symbols represent data taken from ref.\onlinecite{moser:transverse:1998} to show the likely regions of CL and SP. With increased pressure, the nesting of the Fermi surface is broken and the SDW phase is suppressed. $T_s$ drops drastically at around $P$ = 3 GPa and disappears at $P$ = 4.37 GPa. Above $P$ = 4.37 GPa, metallic behaviour is observed over the complete temperature range. The highest $T_c$ was observed near the pressure at which the SDW phase vanished. The SC phase in the vicinity of the SDW phase was observed at 4.18 $\le$ P $\le$ 6.03 GPa. The emergence of superconductivity which competes with the SDW state at 4.18 $\le$ P $\le$ 4.37 GPa appears to be a general behaviour among (TMT$C$F)$_2$X superconductors. The origin of SC of this group is considered to be produced by spin-fluctuation mediated Cooper pairs as discussed among heavy fermion systems since antiferro-magnetic spin fluctuation is developed in the normal state. Our data also supports that the spin fluctuation in the SDW state is enhanced by the formation of SC state. TMTTF salts have higher $T_c$ and higher $P_c$ than TMTSF salts. The higher $T_c$ is associated with the smaller band width due to the smaller transfer integrals, which increases density of states. This feature appears as the anion size dependency in $\beta$-BEDT-TTF group.\cite{williams:exotic:1986}

\subsection{Field Dependence}
There are two independent mechanisms for suppressing superconductivity with magnetic fields; one is orbital breaking of cooper pairs in the superconducting state associated with screening currents generated to exclude the external field (orbital limit) which becomes dominant first for most superconductors. The other mechanism is a spin effect due to Zeeman splitting which applies only to singlet pairings and limits superconductivity below the Pauli or Clongston-Chandrasekhar limit\cite{clogston:1962:upper,chandrasekhar:notemaximum:1962}, as given by weak-coupling BCS paramagnetic formula, 
\begin{eqnarray}
H_p=\frac{\Delta_0}{\sqrt{2}\mu_B}\simeq 1.84T_c(T)
\label{pauli}
\end{eqnarray}
where $\Delta_0$ is the superconducting gap at $T$ = 0 K and $\mu_B$ is Bohr magneton for isotropic $s$-wave pairing in the absence of spin-orbit scattering. This relation is a good guide to determine whether the system is a triplet superconductor or not. The anisotropy of $H_{c2}$ in (TMTSF)$_2$PF$_6$ was determined by Lee\cite{lee:unconventional:2006}. It was more than four times the Pauli limit and there was no change in the NMR night shift. These results supported the view that this system is a triplet superconductor. In the case of (TMTSF)$_2$ClO$_4$, Yonezawa et al.\cite{yonezawa:magnetic-field:2008} conclude that it is a singlet superconductor and exhibits an FFLO state in high fields. Aizawa et al. showed that consecutive transitions from singlet pairing to FFLO and further to $Sz$ = 1 triplet pairing can generally take place upon increasing the magnetic field in the vicinity of the SDW+CDW coexisting phase, and they postulate (TMTSF)$_2$X as a candidate material which exhibits this property.\cite{aizawa:strong:2009}

\begin{figure}
   \includegraphics[width=3in]{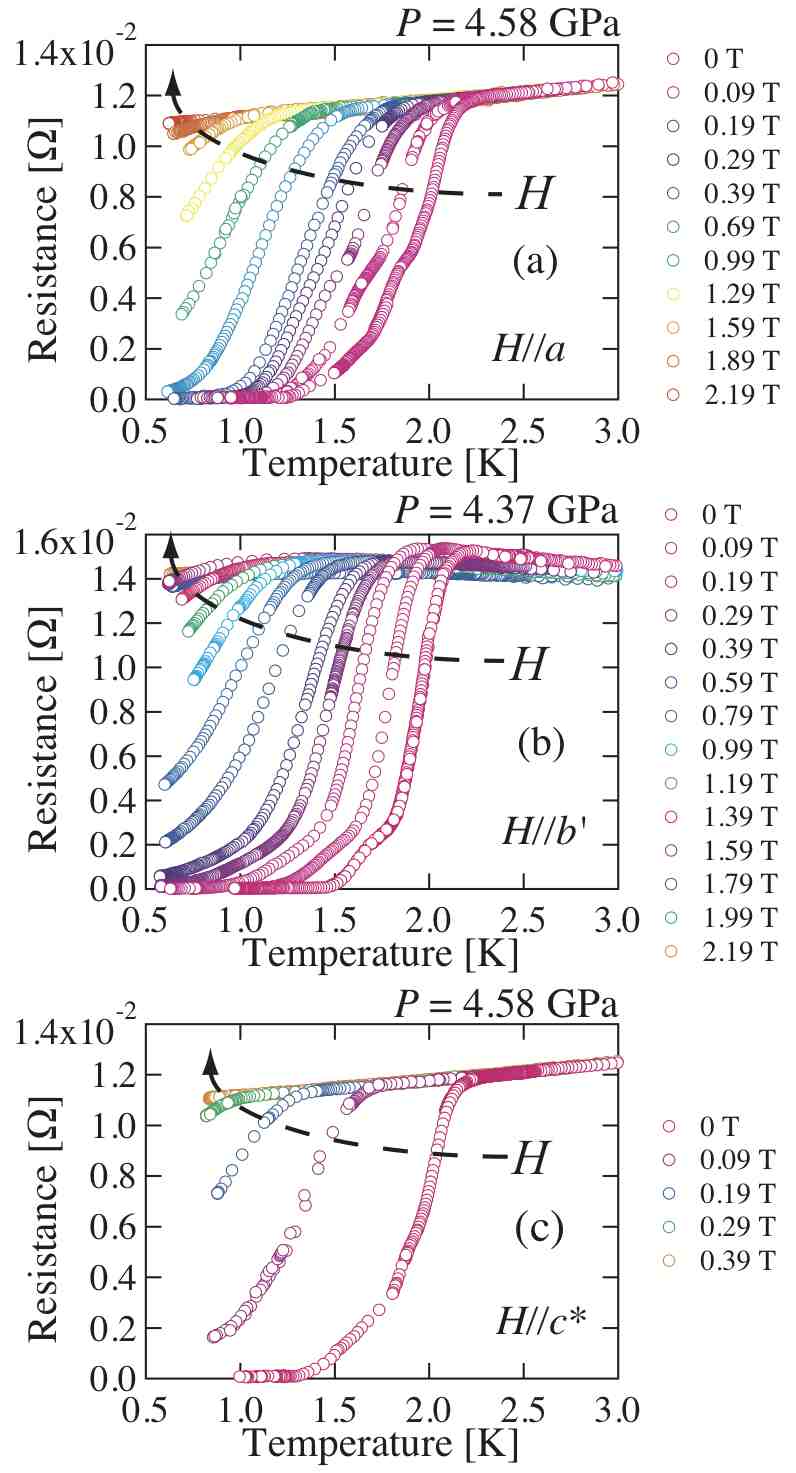} 
   \caption{\label{fig4}Temperature dependent resistance in the magnetic field. Applied magnetic fields were (a) parallel to $a$-axis at $P$ = 4.58 GPa, (b) parallel to $b'$-axis at $P$ = 4.37 GPa, and (c) parallel to $c^*$-axis at $P$ = 4.58 GPa}
\end{figure}

\begin{figure}
   \includegraphics[width=3in]{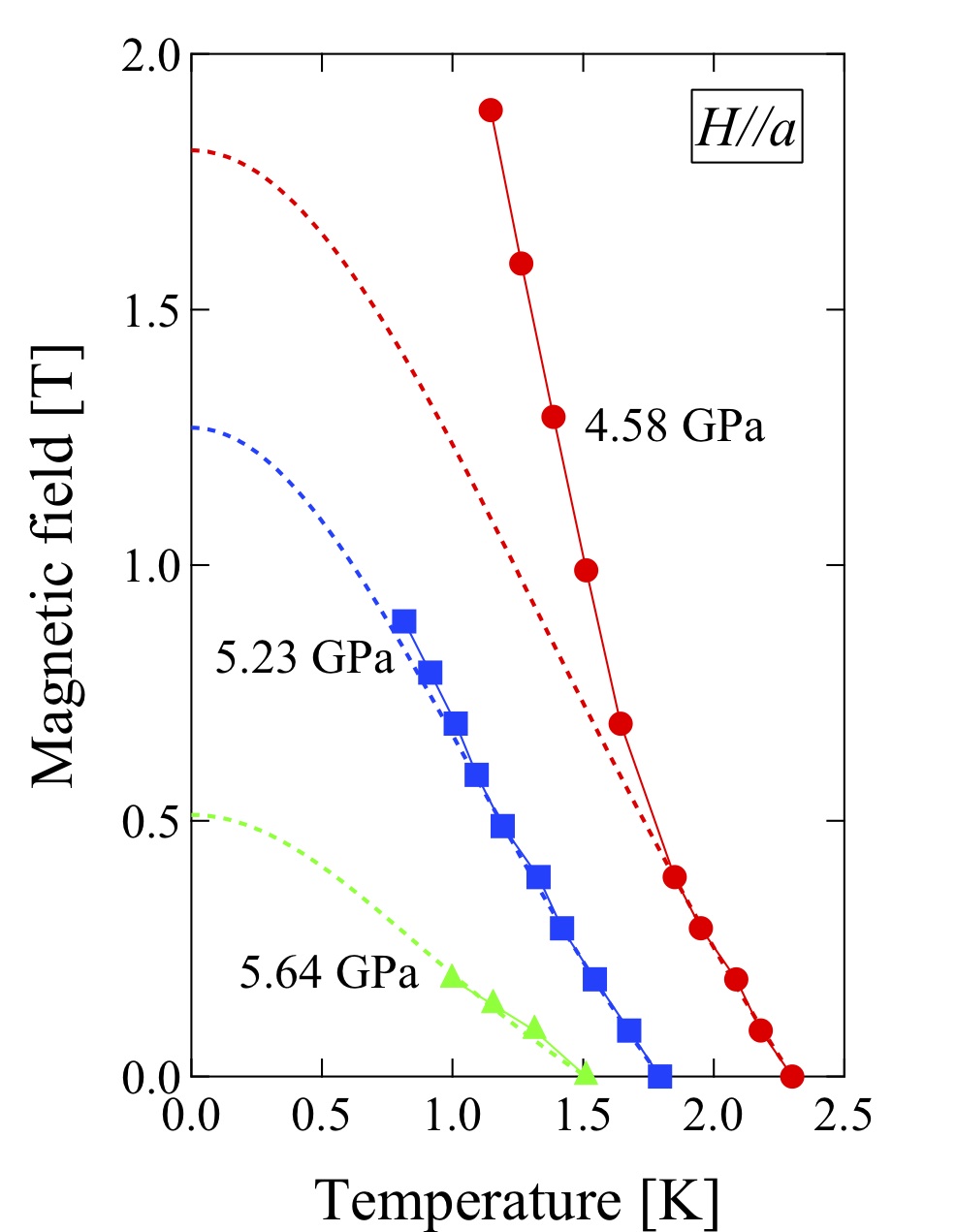} 
   \caption{\label{fig5}Temperature dependence of $H_{c2}$ (resistive onset) at $P$ = 4.58, 5.23 and 5.64 GPa for $H//a$. Dashed lines are GL theoretical fits to show the orbital limiting boundary.}
\end{figure}

\begin{figure}[htbp]
   \includegraphics[width=2.5in]{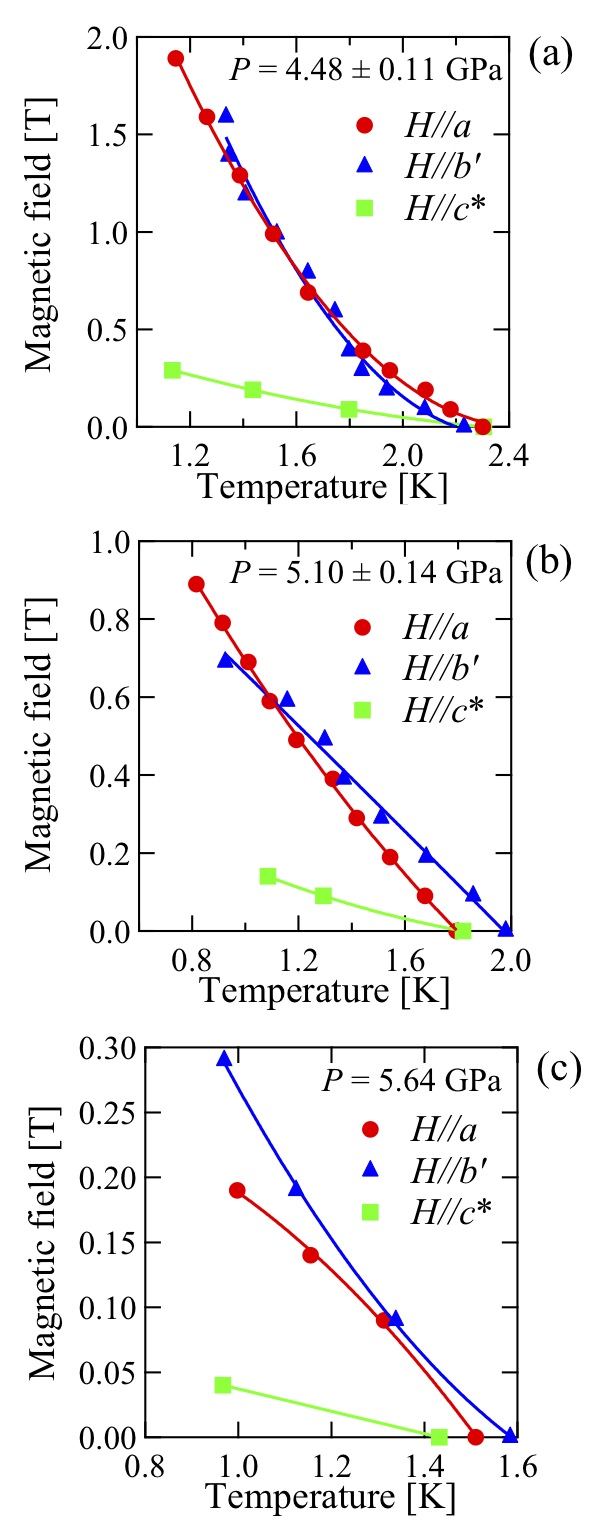} 
   \caption{\label{fig6}$H_{c2}$ curves with the applied field $H//a$, $H//b'$, and $H//c^*$ for $P$ = (a) 4.48, (b) 5.10 and (c) 5.64 GPa obtained by using the onset criterion.}
\end{figure}

We studied how the superconducting transitions are affected by magnetic fields to get detailed information on $H_{c2}$ and its anisotropy. Data of resistivity vs. temperature at $P$ = 4.37 - 4.58 GPa and with the field along three different axes are displayed in FIG.\ref{fig4}. 
For $H//a$ and $c^*$,  the  normal state in zero applied field doesn't change at all as increasing magnetic field and the transition temperatures are suppressed to lower temperature. For $H//b'$, the difference in the values of resistivity is due to the pressure change during the thermal cycling. The pressure rose from 4.37 GPa to 4.58 GPa during the measurement for $H//b'$.  The superconductivity was not observed above 2.2 T for $a$ and $b'$-axis whereas very low field, $\mu_0H$ = 0.39 T for the $c^*$-axis for measured temperature range. FIG.\ref{fig5} is the temperature dependence of $H_{c2}$ (resistive onset) at $P$ = 4.58, 5.23 and 5.64 GPa for $H//a$. 
The slopes of $H_{c2}(T)$ near $T_c$, i.e. $dH_{c2}/dT$, were 0.86, 0.77 and 0.37 TK$^{-1}$ for $P$ = 4.58, 5.23 and 5.64 GPa, respectively.  By using the (WHH) formula\cite{werthamer:temperature:1966}, the values of the zero-temperature upper critical field $H_{c2}(0)$ can be estimated by
\begin{eqnarray}
\mu_0H_{c2}(T = 0) = -0.69T_c\Bigl|\frac{\mu_0\partial H_{c2}}{\partial T}\Bigr|_{T_c}
\label{whh}
\end{eqnarray}
Taking $T_c$ = 2.30, 1.79, 1.51 K, we obtain 1.36, 1.01 and 0.47 T for corresponding pressures. In the Ginzburg-Landau theory, it is known that $H_{c2} = \Phi_0/2\pi\xi^2$ and $\xi\propto\sqrt{(1+t^2)/(1-t^2)}$ where $\Phi_0$ is the flux quantum, $\xi$ is the coherence length, and $t = T/T_c$ is the reduced temperature, thus one has
\begin{eqnarray}
H_{c2}(T)=H_{c2}(0)\frac{1-t^2}{1+t^2}.
\label{glfit}
\end{eqnarray}
We use this equation to fit our data to enhance the deviation from the orbital limiting boundary and show them in FIG.\ref{fig5} as the dashed lines. Slightly higher $H_{c2}(0)$ were achieved by the GL theoretical fit (Eq.\ref{glfit}) than WHH prediction (Eq.\ref{whh}). The estimation from both WHH and GL theory, for $a$-axis, at 4.58 GPa, clearly underestimates $H_{c2}(0)$ obtained from linear extrapolation which is roughly $H(0)\sim$ 4.5 T,  by factor of 2.5 - 3. With increasing pressure, $H_{c2}(T)$ is described better with the GL curve; it loses the upturn feature at low temperatures. This fact indicates that at higher pressure, $H_{c2}(T)$ is mainly determined by the orbital effect and the spin Pauli paramagnetic effect and other mechanism do not play a dominant role.  The Pauli limit for $P$ = 4.58 GPa, $T_c$ = 2.30 K is $H_P\sim$ 4.2 T (Eq.\ref{pauli}), about the same but slightly higher value obtained from low temperature extrapolation, $H(0)\sim$ 4.5 T, suggesting that it is within the range where the paramagnetic pair breaking effect dominates and it is a singlet $s$-wave pairing superconductor at this pressure as it doesn't exceed Pauli limit. Therefore the striking upturn that appeared in the $H_{c2}(T)$ curve at 4.58 T below 1.85 K may come from the FFLO state and so might the slight upturn at 5.23 GPa. Lebed estimated the FFLO field at zero temperature to be $H_P^{FFLO}(0)\simeq 0.6\sqrt{t_a/t_b}H_P$ where $t_a$ and $t_b$ are transfer integrals along $a$ and $b$. \cite{lebed:revival:1999} If you take a conservative estimate of the ratio $t_a$/$t_b$ = 4, it yields $H_P^{FFLO}(0)\sim 5 T$ which seems to be a reasonable value for the FFLO limiting to occur however with the ratio we obtain from coherence length in this work, $t_a$/$t_b\sim$ 1.5, yielding an FFLO limiting field of 3.1 T which is 30\% smaller than the low temperature extrapolated value, $H(0)\sim$ 4.5 T. Further study is needed at lower temperature and also in the pressure range where spin density waves and superconductivity coexist\cite{gorkov:nature:2007} to determine whether the values of $H_{c2}$ exceed this limit as $T$ approaches zero or not, and reveal if the same scenario as TMTSF salts applies to the present case.
\begin{table*}
\caption{\label{table1}The slope of $H_{c2}$ curve near $T_c$ ($dH/dT$), coherent length ($\xi$), anisotropy ($\epsilon$), and the ratio of effective mass ( $m_a^*:m_b^*:m_c^*$ ). See text for details. }
\begin{ruledtabular}
\begin{tabular}{ccccccccccc}
$P$[GPa] & $T_c$[K] & $dH_a/dT$[TK$^{-1}$] & $dH_{b'}/dT$[TK$^{-1}$] & $dH_{c*}/dT$[TK$^{-1}$] & $\xi_a$[$\AA$] & $\xi_{b'}$[$\AA$] & $\xi_{c*}$[$\AA$] & $\epsilon_{ac}$ & $\epsilon_{bc}$ & $m^*_a:m^*_{b'}:m^*_{c*}$\\
\hline
4.48$\pm$0.11&2.27$\pm$0.04&0.864&0.738&0.246&264&223&75.1&3.5&3.0&1.0 : 1.4 : 12\\
5.10$\pm$0.14&1.89$\pm$0.09&0.773&0.630&0.187&338&276&81.8&4.1&3.4&1.0 : 1.5 : 17\\
5.64&1.51$\pm$0.08&0.457&0.465&0.086&499&507&94.0&5.3&5.4&1.0 : 1.0 : 29\\
\end{tabular}
\end{ruledtabular}
\end{table*}
The anisotropy of $H_{c2}$ for $P$ = 4.48, 5.10 and 5.64 GPa obtained by using the onset criterion is shown in FIG.\ref{fig6}. 
Due to the difficulty of achieving exactly the same pressure at low temperatures for each thermal cycle, the plots for all axes that show similar $T_c$ with zero field are compared in the same figure even though the pressures are slightly different. The pressure difference is smaller than 3\%.  A positive curvature without saturation as $T$ approaches low temperature can be seen for $H // a$ and $H // b'$ and they are almost on top of each other at $P$ = 4.48 GPa in this temperature range. Interestingly, similar diagrams were obtained from the measurement in (TMTSF)$_2$PF$_6$. \cite{lee:unconventional:2006} When the field is applied along the $c^*$-axis, the orbital limit dominates because the field penetrates $a-b'$ plane and de-couples the Cooper pairs. However it is unusual that the curvature for $H//b'$ is on top of $H//a$, not showing the anisotropy at ambient pressure at all where the transfer integral $t_b$ is smaller than $t_a$ by one order of magnitude. One may consider from this behavior that paramagnetic limiting dominates for both orientations, but as can be seen in FIG.\ref{fig6}(b) and  (c), the same curvatures appear at higher pressures where the Pauli limit is far above and the orbital limiting clearly dominates. Therefore this feature seems universal among TMT$C$F salts.

A characteristic value of superconductors is the coherence length, $\xi$. $\xi$ is the spatial parameter for the wave function of the Cooper pair, and electrons within this length move while correlating with each other. We estimated the Ginzburg-Landau (GL) coherence lengths from the slope of $H_{c2}$ for three directions near $T_c$ by using the following relations (Eq.\ref{xi}) for anisotropic superconductors (Table.\ref{table1}).  
\begin{eqnarray}
\Bigl|\frac{\partial H_{c2}^i}{\partial T}\Bigr|_{T_c}=\frac{\Phi_0}{2\pi\xi_k\xi_jT_c}
\label{xi}
\end{eqnarray}
where $\phi_0$ is a flux quantum. $\xi$ values increase monotonically with pressure. However, the coherence lengths are highly anisotropic, for example at $P$ = 5.64 GPa, $\xi_a:\xi_{b'}:\xi_{c*}\sim$ 5 : 5 : 1, the interlayer coherence length, $\xi_{c^*}$, is much longer than the thickness of the conducting sheet which is $c/2\sim$ 6.5 $\AA$ where $c$ is a lattice parameter of (TMTTF)$_2$PF$_6$ for all pressures. These results indicate that the present compound is an anisotropic three-dimensional superconductor which doesn't involve Josephoson interaction. Note that these values might be underestimated due to contribution of the paramagnetic limiting effect leading $H_{c2}$ to have a smaller slope. Therefore the coherence length for $P$ = 4.48 GPa could be slightly smaller if the paramagnetic limiting is completely eliminated but it should not affect this estimation. As seen in FIG.\ref{fig5} for 4.48 GPa, the upturn behaviour in $H_{c2}$ towards low temperature is often considered as the field induced dimensional cross over (FIDC) from 3D to 2D but the calculated $\xi_c$ value shows that there is no FIDC. In the anisotropic 3D model, GL coherence lengths are associated with effective masses, and anisotropy $\epsilon$ is defined as follows.
\begin{eqnarray}
\epsilon_{\parallel\perp}=\frac{H_{c2_\parallel}}{H_{c2_\perp}}=\frac{\xi_\parallel}{\xi_\perp}=\sqrt{\frac{m^*_\perp}{m^*_\parallel}}.
\label{anisotropy}
\end{eqnarray}
We used $\epsilon_{ac}=\xi_a/\xi_c$ and $\epsilon_{bc}=\xi_b/\xi_c$ for our case, and obtained the anisotropy and the ratio of effective masses (Table.\ref{table1}). The ratio of transfer integrals along each axis are estimated as $t_a : t_b : t_c$ = 300 : 10 : 1 at ambient pressure and  room temperature whereas our data, within the pressure and temperature range superconductivity is observed, the strong one dimensional feature is suppressed and dimensionality is increased. The conductivities along $a$ and $b$-axis are about the same and larger than that of the $c$-axis by factor of 10 - 30. With increasing pressure, $m_c$ increases. This is consistent with the results of Morosin et al.\cite{morosin:compressibilities:1982}, that the compressibility is smaller along the $c^*$-axis where a TMTTF molecule and anion have ion-bonding than the $a$-axis where $\pi$ orbit of TMTTF are overlapped with van der Waals bonding.   

\section{Conclusions}
We have conducted transport measurements by using a turnbuckle type DAC on (TMTTF)$_2$PF$_6$ under multi-extreme conditions which had been difficult to achieve at the same time. (TMTTF)$_2$X has higher critical pressure which exceeds the range of widely-used piston cylinder cells, and for this reason not many observations of superconductivity have been reported. No similar measurements, revealing angular dependance of $H_{c2}$ by conducting detailed transport measurements in magnetic field, have been reported prior to this work. We have observed superconductivity in (TMTTF)$_2$PF$_6$ with actual zero resistivity at $P$ = 4.18 - 4.58 GPa. Zero resistivity was never observed in SC phase in other pressure work using a DAC or Bridgman pressure apparatus. Only measurements with a Cubic Anvil Press which generates highly hydrostatic pressures showed superconductivity with zero resistivity\cite{araki:electrical:2007} indicating that our measurement with a DAC has similar accuracy to Cubic Anvil Press. The SC dome lays at $P$ = 4.18 - 6.18 GPa in the $P-T$ phase diagram obtained from this work and the highest transition temperature was $T_c$ = 2.25 K. Our data extended the range of temperature and pressure and supplemented the reported $P-T$ phase diagram.

Also we have shown the temperature dependence of $H_{c2}$ of three axis on (TMTTF)$_2$PF$_6$ and that $H_{c2}$ displays positive curvature without saturation which may be attributed to an FFLO state, for magnetic field along $a$-axis and $b'$-axis in T$\ge$0.5 K for P = 4.48 GPa. We also have shown that the upturn feature is suppressed with increasing pressure and the orbital pair breaking mechanism becomes dominant. In further studies on the pair breaking mechanism, measurements need to be conducted at lower temperatures and with more refined pressure control. Our data above $T$ = 0.5 K shows $H_{c2}$ slightly exceed the Pauli paramagnetic limit at $P$ = 4.48 GPa for $H//a$. $H_{c2}$ curves for $H//a$ and $H//b'$ lie on top of each other indicating that it has two-dimentional-like feature being isotropic within an $a-b'$ plane. GL coherence lengths were obtained and revealed that (TMTTF)$_2$PF$_6$ is anisotropic three dimensional superconductor. 

To estimate the relationship of anion size and the symmetry of SC, family compounds such as (TMTTF)$_2$SbF$_6$, AsF$_6$ are necessary. Pressure is the easiest and the strongest tool to manipulate properties of the sample. As discussed in introduction, we need concrete information on how the pressure changes the distance between molecules for each axis and affects electronic correlation to reveal the difference between the chemical and external pressures.

\begin{acknowledgments}
We wish to thank Dr. Y. Huseya for useful discussions. This work was supported by Technology (MEXT) of the Japanese Government, Grant-in-Aid For Scientific Research (21340092, 21740263, and 21110517) of JSPS of Japan. This work has been performed at the Institute for Solid State Physics, University of Tokyo.
\end{acknowledgments}

\bibliography{tmttf}

\end{document}